\newcommand{\PLA}[3]{Phys.\ Lett.\ A\ {\bf #1},\ #2 (#3)}

\newcommand{\PRL}[3]{Phys.\ Rev.\ Lett.\ {\bf #1},\ #2 (#3)}

\newcommand{\NAT}[3]{Nature\ {\bf #1},\ #2 (#3)}

\newcommand{\PRA}[3]{Phys.\ Rev.\ A\ {\bf #1},\ #2 (#3)}

\newcommand{\JPA}[3]{J.\ Phys.\ A:\ Math.\ Gen.\ {\bf #1},\ #2 (#3)}

\documentclass[aps,showkeys,showpacs,amsmath,amssymb]{revtex4}
\usepackage{dcolumn}
\usepackage{longtable}

\begin{document}
\title{Perfect teleportation and superdense coding through an asymmetric five qubit state}
\author{Sreraman Muralidharan}
\email{sreraman@loyolacollege.edu} \affiliation{Loyola College,
Nungambakkam, Chennai - 600 034, India}
\author{Prasanta K. Panigrahi}
\email{prasanta@prl.res.in} 
\affiliation{Indian Institute of Science Education and Research (IISER) Kolkata, Salt Lake, Kolkata - 700106, India}
\affiliation{Physical Research
Laboratory, Navrangpura, Ahmedabad - 380 009, India}

\begin{abstract}
We explicate an example of the so called Task oriented Maximally Entangled states (TMES) in the context
of teleportation and superdense coding. In physical situation, this
state can emerge from decoherence of more entangled state or may be prepared for the 
purpose at hand, with less entanglement resource. We find that a variant of the Brown
state can be utilized for perfect teleportation of single and two qubit states. In case of 
superdense coding, it is observed that five cbits can be transmitted by sending only
three qubits.
\end{abstract}



\maketitle

\section{Introduction}
Entanglement is a completely quantum characteristic which has given rise to a
number of counter-intuitive phenomena. The most well known of this is the 
Einstein-Podolsky-Rosen (EPR) paradox, which arises because of quantum
correlations. With the advent of quantum computation, this correlation
has come in handy for a number of purposes, not achievable in the classical
world. In a remarkable result \cite{Bennett}, it has been shown that an unknown single
qubit state can be teleported to a desired destination through an EPR state.
Entanglement is well understood in the case of two particles. In the multi-particle
scenario, much remains to be understood and explored. In the case of three particles,
the well known states are the GHZ and the W states. Interestingly, the former can
be used for teleportation but the latter is unusable for the same purpose. In the case of four particles, 
entangled Bell pair and the cluster states have found much applications
in Quantum computation. These states have been experimentally
realized in laboratory conditions \cite{china}. Higher particle generalizations are of recent origin.
Prominent among them is the five particle Brown state \cite{Brown}, which has
been numerically shown to be highly entangled. Recently, the utility of this
state for teleportation, Quantum state sharing and superdense coding have been
illustrated by us \cite{Sreraman}. These kind of states for the purpose of
specific tasks are called Task oriented Maximally entangled states (TMES) \cite{Pankaj}.  

Superdense coding is intimately related to teleportation, whereas in the 
case of state-splitting , multiple parties need to cooperate for a given
member to recover the state. In reality, entangled states are prone
to decoherence. In many scenarios, highly entangled states may not be required
for the purpose at hand. Hence, it is worth investigating a less entangled
state for physical applications. Below, we consider a variant of the Brown state
for usefulness towards teleportation and dense coding. Also, the state is highly
robust from the point of view of entanglement. This state is asymmetric in the sense 
that it does not have the same form for all $(3+2)$ splits.  
This state has the form :
\begin{equation}
|\psi_{5}\rangle=\frac{1}{2}(|000\rangle|\phi_{-}\rangle+|010\rangle|\psi_{-}\rangle
+|100\rangle|\phi_{+}\rangle+|111\rangle|\psi_{+}\rangle),
\end{equation}

where, $|\psi_\pm\rangle=|00\rangle\pm|11\rangle$ and
$|\phi_\pm\rangle=|01\rangle\pm|10\rangle$ are Bell states. $|\psi_5\rangle$ 
exhibits genuine multi-partite entanglement according to
both negative partial transpose measure, as well as von Neumann
entropy measure. The von Neumann entropy between (1234$|$5) is
equal to one and between (123$|$45) is two. These are the maximum
possible entanglement values between the respective subsets. Even
after tracing out one/two qubits from the state, entanglement
sustains in the resulting subsystem and thus, is highly `robust'.
Also, the state is maximally mixed, after we trace out $1$, or
$2$, or $3$ or $4$ qubits which is an indication of genuine
multi-particle entanglement for the five-qubit state
$|\psi_5\rangle$. Four-qubit states do not show such
characteristic behaviour and fail to attain maximal entropy
\cite{Higuchi}. Thus, the five-qubit state can provide an edge
over the four-qubit state for state transfer and coding.
We recall the standard protocol for teleportation for the sake of
completeness. Alice and Bob share a maximally entangled state
$|\varphi\rangle_{AB}$, where A and B respectively refer to the
subsystems of Alice and Bob. Alice wants to teleport
$|\psi\rangle_a$, in her possession, to Bob. Thus, she prepares
the combined state,

\begin{equation}
|\psi\rangle_{a}|\varphi\rangle_{AB}= \frac{1}{\sqrt{D}}
\sum_{x=1}^{D}|\phi_{x}\rangle_{aA}U_{x}|\psi\rangle_{B}.
\end{equation}

Here, ${U_x}$ are unitary operators on subsystem B and
${|\phi_{x}\rangle_{aA}}$ are mutually orthogonal states of the
joint system. Now, Alice makes a projective measurement on the
joint system and classically communicates the result to Bob, who
then recovers the state, after applying appropriate unitary
operations.

\section{A single qubit state}
In the seminal paper, Bennett \it{et al.} \normalfont \cite{Bennett}  considered the 
teleportation of a single qubit state given by $\alpha|0\rangle+\beta|1\rangle$
using the Bell state as an entangled channel. In their scheme,
Alice first combines the unknown qubit with her state and 
peforms a Bell measurment and communicates the result
of her measurement to Bob via two cbits of information. 
Bob, then performs an unitary operation and obtains the 
unknown qubit. We now demonstrate, the utility of $|\psi_{5}\rangle$
for teleportation. Let us first consider the situation in which Alice possesses
qubits $1$, $2$, $3$, $4$ and particle $5$ belongs to Bob. Alice
wants to teleport $\alpha|0\rangle+\beta|1\rangle$ to Bob. So,
Alice prepares the combined state,

\begin{eqnarray}
(\alpha|0\rangle+\beta|1\rangle)|\psi_{5}\rangle=|\phi_{1}\rangle_{a_{1+}}
(\alpha|0\rangle+\beta|1\rangle)+|\phi_{2}\rangle_{a_{1-}}\alpha|0\rangle-\beta|1\rangle)
+|\phi_{3}\rangle_{a_{2+}}(\beta|0\rangle+\alpha|1\rangle)+|\phi_{4}\rangle_{a_{2-}}(\beta|0\rangle-\alpha|1\rangle),
\end{eqnarray}

where, the $|\phi_{x}\rangle_{a_{i}\pm}$ are mutually orthogonal
states of the measurement basis. The states
$|\phi_{x}\rangle_{a_{i}\pm}$ are given as,

\begin{eqnarray}
|\phi_{x}\rangle_{a_{1}\pm}&=&(-|00001\rangle_{A}+|00100\rangle_{A}+|01001\rangle_{A}+|01110\rangle_{A}\pm(|10000\rangle_{A}-
|10101\rangle_{A}+|11000\rangle_{A}\nonumber+|11111\rangle_{A}),\\
|\phi_{x}\rangle_{a_{2}\pm}&=&(-|10001\rangle_{A}+|10100\rangle_{A}+|11001\rangle_{A}+|11110\rangle)_{A}\pm(|00000\rangle_{A}-
|00101\rangle_{A}+|01000\rangle_{A}\nonumber+|01111\rangle_{A}).
\end{eqnarray}

Alice can now make a five-particle measurement using
$|\phi_{x}\rangle_{a_{i}\pm}$. Bob can apply suitable unitary
operations given by $(1,\sigma_{1},i\sigma_{2},\sigma_{3})$ to
recover the original state $(\alpha|0\rangle+\beta|1\rangle)$.
This completes the teleportation protocol for the teleportation of
a single qubit state using the state $|\psi_5\rangle$.

\section{An arbitrary two qubit state}
Teleportation of an arbitrary two qubit state was first studied, by Rigolin \cite{Rigolin1},
using two entangled Bell pairs as a quantum channel. Later, a genuinely entangled four qubit state
was introduced by Yeo and Chua \cite{Yeo} for the same purpose. In this section, we shall
investigate the efficacy of $|\psi_{5}\rangle$ for the same purpose. 
Alice has an arbitrary two qubit state,

\begin{equation}
|\psi\rangle=\alpha|00\rangle+\mu|10\rangle+\gamma|01\rangle+\beta|11\rangle,
\end{equation}

which she has to teleport to Bob. Qubits $1$, $2$, $3$ and $4$,
$5$ respectively, belong to Alice and Bob. Alice prepares the
combined state,

\begin{eqnarray}
|\psi\rangle|\psi_{5}\rangle & = &
\frac{1}{4}[|\psi_{5}\rangle_1(\alpha|01\rangle+\gamma|00\rangle+\mu|11\rangle+\beta|10\rangle)+
|\psi_5\rangle_2(\alpha|01\rangle+\gamma|00\rangle-\mu|11\rangle-\beta|10\rangle)+\nonumber
\\ & &
|\psi_5\rangle_3(\alpha|01\rangle-\gamma|00\rangle+\mu|11\rangle-\beta|10\rangle)+
|\psi_5\rangle_4(\alpha|01\rangle-\gamma|00\rangle-\mu|11\rangle+\beta|10\rangle)+\nonumber
\\ & &
|\psi_5\rangle_5(\alpha|11\rangle+\gamma|10\rangle+\mu|01\rangle+\beta|00\rangle)+
|\psi_5\rangle_6(\alpha|11\rangle-\gamma|10\rangle+\mu|01\rangle-\beta|00\rangle)+\nonumber
\\ & &
|\psi_5\rangle_7(\alpha|11\rangle+\gamma|10\rangle-\mu|01\rangle-\beta|00\rangle)+
|\psi_5\rangle_8(\alpha|11\rangle-\gamma|10\rangle-\mu|01\rangle+\beta|00\rangle)+\nonumber
\\ & &
|\psi_5\rangle_9(\alpha|00\rangle+\gamma|01\rangle+\mu|10\rangle+\beta|11\rangle)+
|\psi_5\rangle_{10}(\alpha|00\rangle-\gamma|01\rangle+\mu|10\rangle-\beta|11\rangle)+\nonumber
\\ & &
|\psi_5\rangle_{11}(\alpha|00\rangle+\gamma|01\rangle-\mu|10\rangle-\beta|11\rangle)+
|\psi_5\rangle_{12}(\alpha|00\rangle-\gamma|01\rangle-\mu|10\rangle+\beta|11\rangle)+\nonumber
\\ & &
|\psi_5\rangle_{13}(\alpha|10\rangle+\gamma|11\rangle+\mu|00\rangle+\beta|01\rangle)+
|\psi_5\rangle_{14}(\alpha|10\rangle-\gamma|11\rangle+\mu|00\rangle-\beta|01\rangle)+\nonumber
\\ & &
|\psi_5\rangle_{15}(\alpha|10\rangle+\gamma|11\rangle-\mu|00\rangle-\beta|01\rangle)+
|\psi_5\rangle_{16}(\alpha|10\rangle-\gamma|11\rangle-\mu|00\rangle+\beta|01\rangle)].
\end{eqnarray}

Here, $|\psi_5\rangle_{i}$'s forming the mutual orthogonal basis
of measurement are given by :

\begin{eqnarray}
|\psi_{5}\rangle_{1}=\frac{1}{2}[|\phi_{-}\rangle|010\rangle+|\phi_{+}\rangle|111\rangle+|\psi_{-}\rangle|000\rangle+|\psi_{+}\rangle|
100\rangle];
|\psi_{5}\rangle_{2}=\frac{1}{2}[|\phi_{+}\rangle|010\rangle+|\phi_{-}\rangle|111\rangle+|\psi_{+}\rangle|000\rangle+|\psi_{-}\rangle|
100\rangle];\nonumber \\
|\psi_{5}\rangle_{3}=\frac{1}{2}[|\psi_{-}\rangle|000\rangle+|\psi_{+}\rangle|100\rangle-|\phi_{+}\rangle|010\rangle-|\phi_{-}\rangle|
111\rangle];
|\psi_{5}\rangle_{4}=\frac{1}{2}[|\psi_{-}\rangle|000\rangle+|\psi_{+}\rangle|100\rangle-|\phi_{-}\rangle|010\rangle-|\phi_{+}\rangle|
111\rangle];\nonumber \\
|\psi_{5}\rangle_{5}=\frac{1}{2}[|\psi_{+}\rangle|111\rangle-|\psi_{-}\rangle|010\rangle-|\phi_{-}\rangle|000\rangle+|\phi_{+}\rangle|
100\rangle];
|\psi_{5}\rangle_{6}=\frac{1}{2}[|\psi_{-}\rangle|111\rangle-|\psi_{+}\rangle|010\rangle+|\phi_{+}\rangle|000\rangle-|\phi_{-}\rangle|
100\rangle];\nonumber \\
|\psi_{5}\rangle_{7}=\frac{1}{2}[|\psi_{-}\rangle|111\rangle-|\psi_{+}\rangle|010\rangle-|\phi_{+}\rangle|000\rangle+|\phi_{-}\rangle|
100\rangle];
|\psi_{5}\rangle_{8}=\frac{1}{2}[|\psi_{+}\rangle|111\rangle-|\psi_{-}\rangle|010\rangle+|\phi_{-}\rangle|000\rangle-|\phi_{+}\rangle|
100\rangle];\nonumber \\
|\psi_{5}\rangle_{9}=\frac{1}{2}[|\psi_{+}\rangle|111\rangle+|\psi_{-}\rangle|010\rangle+|\phi_{-}\rangle|000\rangle+|\phi_{+}\rangle|
000\rangle];
|\psi_{5}\rangle_{10}=\frac{1}{2}[|\psi_{-}\rangle|111\rangle+|\psi_{+}\rangle|010\rangle-|\phi_{-}\rangle|000\rangle-|\psi_{+}\rangle|
100\rangle];\nonumber \\
|\psi_{5}\rangle_{11}=\frac{1}{2}[|\psi_{-}\rangle|111\rangle+|\psi_{+}\rangle|010\rangle+|\phi_{+}\rangle|000\rangle+|\phi_{-}\rangle|
100\rangle];
|\psi_{5}\rangle_{12}=\frac{1}{2}[|\psi_{+}\rangle|111\rangle+|\psi_{-}\rangle|010\rangle-|\phi_{+}\rangle|100\rangle-|\phi_{-}\rangle|
000\rangle];\nonumber \\
|\psi_{5}\rangle_{13}=\frac{1}{2}[|\psi_{+}\rangle|100\rangle-|\psi_{-}\rangle|000\rangle-|\phi_{-}\rangle|010\rangle+|\phi_{+}\rangle|
111\rangle];
|\psi_{5}\rangle_{14}=\frac{1}{2}[|\psi_{-}\rangle|100\rangle-|\psi_{+}\rangle|000\rangle+|\phi_{+}\rangle|010\rangle-|\phi_{-}\rangle|
111\rangle];\nonumber \\
|\psi_{5}\rangle_{15}=\frac{1}{2}[|\psi_{-}\rangle|100\rangle-|\psi_{+}\rangle|000\rangle-|\phi_{+}\rangle|010\rangle+|\phi_{-}\rangle|
111\rangle];
|\psi_{5}\rangle_{16}=\frac{1}{2}[|\psi_{+}\rangle|100\rangle-|\psi_{-}\rangle|000\rangle+|\phi_{-}\rangle|010\rangle-|\phi_{+}\rangle|
111\rangle].
\end{eqnarray}

Alice can make a five-particle measurement and then convey her
results to Bob. Bob then retrieves the original state
$|\psi\rangle_{b}$ by applying any one of the unitary transforms
shown in Table \ref{tab1} to the respective states. As is evident,
each of the above states are obtained with equal probability. This
successfully completes the teleportation protocol of a two qubit
state using $|\psi_{5}\rangle$.

\begin{table}[h]
\caption{\label{tab1}Set of Unitary operators needed to obtain
$|\psi\rangle_b$}
\begin{tabular}{|c|c|}
\hline {\bf State} & {\bf Unitary}\\
& {\bf Operation}\\
\hline
$(\alpha|01\rangle+\gamma|00\rangle+\mu|11\rangle+\beta|10\rangle)$
& $I\otimes\sigma_{1}$\\
$(\alpha|01\rangle+\gamma|00\rangle-\mu|11\rangle-\beta|10\rangle)$
& $\sigma_{3}\otimes\sigma_{1}$\\
$(\alpha|01\rangle-\gamma|00\rangle+\mu|11\rangle-\beta|10\rangle)$&
$I\otimes i\sigma_{2}$\\
$(\alpha|01\rangle-\gamma|00\rangle-\mu|11\rangle+\beta|10\rangle)$&
$\sigma_{3}\otimes i\sigma_{2}$\\
$(\alpha|11\rangle+\gamma|10\rangle+\mu|01\rangle+\beta|00\rangle)$
& $\sigma_{1}\otimes\sigma_{1}$\\
$(\alpha|11\rangle-\gamma|10\rangle+\mu|01\rangle-\beta|00\rangle)$&
$\sigma_{1}\otimes i\sigma_{2}$\\
$(\alpha|11\rangle+\gamma|10\rangle-\mu|01\rangle-\beta|00\rangle)$&
$i\sigma_{2}\otimes\sigma_{1}$\\
$(\alpha|11\rangle-\gamma|10\rangle-\mu|01\rangle+\beta|00\rangle)$
& $i\sigma_{2}\otimes i\sigma_{2}$\\
$(\alpha|00\rangle+\gamma|01\rangle+\mu|10\rangle+\beta|11\rangle)$ & $I\otimes I$\\
$(\alpha|00\rangle-\gamma|01\rangle+\mu|10\rangle-\beta|11\rangle)$ & $I\otimes\sigma_{3}$\\
$(\alpha|00\rangle+\gamma|01\rangle-\mu|10\rangle-\beta|11\rangle)$
& $\sigma_{3}\otimes I$\\
$(\alpha|00\rangle-\gamma|01\rangle-\mu|10\rangle+\beta|11\rangle)$
& $\sigma_{3}\otimes\sigma_{3}$\\
$(\alpha|10\rangle+\gamma|11\rangle+\mu|00\rangle+\beta|01\rangle)$
& $\sigma_{1}\otimes I$\\
$(\alpha|10\rangle-\gamma|11\rangle+\mu|00\rangle-\beta|01\rangle)$
& $\sigma_{1}\otimes\sigma_{3}$\\
$(\alpha|10\rangle+\gamma|11\rangle-\mu|00\rangle-\beta|01\rangle)$
& $i\sigma_{2}\otimes I$\\
$
(\alpha|10\rangle-\gamma|11\rangle-\mu|00\rangle+\beta|01\rangle)$
&
$i\sigma_{2}\otimes\sigma_{3}$\\
\hline
\end{tabular}
\end{table}

\section{Superdense coding}
We now proceed to show the utility of $|\psi_{5}\rangle$ for
superdense coding. Entanglement is quite handy in communicating
information efficiently, in a quantum channel. Suppose Alice and
Bob share an entangled state, namely $|\psi>_{AB}$, then Alice can
convert her state into different orthogonal states by applying
suitable unitary transforms on her particle \cite{Wiesner}. Bob
then does appropriate Bell measurements on his qubits to retrieve
the encoded information. It is known that two classical bits per
qubit can be exchanged by sending information through a Bell
state. In this section, we shall discuss the suitability of
$|\psi_5\rangle$, as a resource for superdense coding. Let us
assume that Alice has first three qubits, and Bob has last two
qubits. Alice can apply the set of unitary transforms on her
particle and generate $64$ states out of which $32$ are mutually
orthogonal as shown below:

\begin{equation}
U^{3}_{x}\otimes I\otimes I\rightarrow|\psi_{5}\rangle_{x_{i}}.
\end{equation}

Bob can then perform a five-partite measurement in the basis of
$|\psi_{5}\rangle_{x_{i}}$ and distinguish these states. The
appropriate unitary transforms applied and the respective states
obtained by Alice are shown in the Table \ref{tab2}.

\begin{table}[h]
\caption{\label{tab2} States $|\psi_{5}\rangle_{x_{i}}$ obtained
by Alice after performing unitary operations $U^{3}_{x}$}
\begin{tabular}{|c|c|}
\hline {\bf Unitary Operation} & {\bf State} \\
\hline$I\otimes I\otimes I$ &
$\frac{1}{2}(|000\rangle|\phi_{-}\rangle+|010\rangle|\psi_{-}\rangle+|100\rangle|\phi_{+}\rangle+|111\rangle
|\psi_{+}\rangle)$\\
$I\otimes\sigma_{3}\otimes I$ &
$\frac{1}{2}(|000\rangle|\phi_{-}\rangle-|010\rangle|\psi_{-}\rangle+|100\rangle|\phi_{+}\rangle-|111\rangle
|\psi_{+}\rangle)$\\
$\sigma_{3}\otimes I\otimes I$ &
$\frac{1}{2}(|000\rangle|\phi_{-}\rangle+|010\rangle|\psi_{-}\rangle-|100\rangle|\phi_{+}\rangle-
|111\rangle|\psi_{+}\rangle)$\\
$\sigma_{3}\otimes \sigma_{3}\otimes I$ &
$\frac{1}{2}(|000\rangle|\phi_{-}\rangle-|010\rangle|\psi_{-}\rangle-|100\rangle|\phi_{+}\rangle
+|111\rangle|\psi_{+}\rangle)$\\
$\sigma_{1}\otimes \sigma_{1}\otimes I$ & $\frac{1}{2}(|110\rangle|\phi_{-}\rangle+|100\rangle|\psi_{-}\rangle+|010\rangle|\phi_{+}\rangle
+|001\rangle|\psi_{+}\rangle)$\\
$\sigma_{1}\otimes i\sigma_{2}\otimes I$& $\frac{1}{2}(|110\rangle|\phi_{-}\rangle-|100\rangle|\psi_{-}\rangle+|010\rangle|\phi_{+}\rangle
-|001\rangle|\psi_{+}\rangle)$\\
$i\sigma_{2}\otimes \sigma_{1}\otimes I$ &
$\frac{1}{2}(|110\rangle|\phi_{-}\rangle+|100\rangle|\psi_{-}\rangle-|010\rangle|\phi_{+}\rangle
-|001\rangle|\psi_{+}\rangle)$\\
$i\sigma_{2}\otimes i\sigma_{2}\otimes I$ &
$\frac{1}{2}(|110\rangle|\phi_{-}\rangle-|100\rangle|\psi_{-}\rangle-|010\rangle|\phi_{+}\rangle
+|001\rangle|\psi_{+}\rangle)$\\
$I\otimes\sigma_{1}\otimes I$ & $\frac{1}{2}(|010\rangle|\phi_{-}\rangle+|000\rangle|\psi_{-}\rangle+|110\rangle|\phi_{+}\rangle
+|101\rangle|\psi_{+}\rangle)$\\
$I\otimes i \sigma_{2}\otimes I$ & $\frac{1}{2}(|010\rangle|\phi_{-}\rangle-|000\rangle|\psi_{-}\rangle+|110\rangle|\phi_{+}\rangle
-|101\rangle|\psi_{+}\rangle)$\\
$\sigma_{3}\otimes\sigma_{1}\otimes I$ & $\frac{1}{2}(|010\rangle|\phi_{-}\rangle+|000\rangle|\psi_{-}\rangle-|110\rangle|\phi_{+}\rangle-
|101\rangle|\psi_{+}\rangle)$\\
$\sigma_{3}\otimes i\sigma_{2}\otimes I$ & $\frac{1}{2}(|001\rangle|\psi_{-}\rangle-|011\rangle|\phi_{-}\rangle-|110\rangle|\phi_{+}\rangle
+|101\rangle|\psi_{+}\rangle)$\\
$\sigma_{1}\otimes I\otimes I$ & $\frac{1}{2}(|100\rangle|\phi_{-}\rangle+|110\rangle|\psi_{-}\rangle+|000\rangle|\phi_{+}\rangle
+|011\rangle|\psi_{+}\rangle) $\\
$\sigma_{1} \otimes \sigma_{3}\otimes I$ & $\frac{1}{2}(|101\rangle|\phi_{-}\rangle-|110\rangle|\psi_{-}\rangle+|000\rangle|\phi_{+}
\rangle-|011\rangle|\psi_{+}\rangle)$\\
$i\sigma_{2}\otimes I\otimes I$ & $\frac{1}{2}(|001\rangle|\phi_{+}\rangle-|101\rangle|\phi_{-}\rangle-|110\rangle|\psi_{-}\rangle
+|011\rangle|\psi_{+}\rangle)$\\
$i\sigma_{2}\otimes\sigma_{3}\otimes I$ & $ \frac{1}{2}(|100\rangle|\phi_{-}\rangle+|110\rangle|\psi_{-}\rangle-|000\rangle|\phi_{+}\rangle
-|011\rangle|\psi_{+}\rangle)$\\
$I\otimes I\otimes \sigma_{1}$ & $\frac{1}{2}(|001\rangle|\phi_{-}\rangle+|011\rangle|\psi_{-}\rangle+|101\rangle|\phi_{+}\rangle
+|110\rangle|\psi_{+}\rangle) $\\
$I\otimes\sigma_{3}\otimes \sigma_{1}$ & $\frac{1}{2}(|001\rangle|\phi_{-}\rangle-|011\rangle|\psi_{-}\rangle+|101\rangle|\phi_{+}\rangle
-|110\rangle|\psi_{+}\rangle)$\\
$\sigma_{3}\otimes \sigma_{1}\otimes \sigma_{1}$ & $\frac{1}{2}(|001\rangle|\phi_{-}\rangle+|011\rangle|\psi_{-}\rangle-|101\rangle
|\phi_{+}\rangle-|110\rangle|\psi_{+}\rangle)$\\
$\sigma_{3}\otimes \sigma_{3}\otimes \sigma_{1}$ & $\frac{1}{2}(|001\rangle|\phi_{-}\rangle-|011\rangle|\psi_{-}\rangle-|101\rangle
|\phi_{+}\rangle+|110\rangle|\psi_{+}\rangle)$\\
$\sigma_{1}\otimes \sigma_{1}\otimes \sigma_{1}$ & $\frac{1}{2}(|111\rangle|\phi_{-}\rangle+|101\rangle|\psi_{-}\rangle+|011\rangle
|\phi_{+}\rangle+|000\rangle|\psi_{+}\rangle)$\\
$\sigma_{1}\otimes i\sigma_{2}\otimes \sigma_{1}$ & $\frac{1}{2}(|111\rangle|\phi_{-}\rangle-|100\rangle|\psi_{-}\rangle+|011\rangle
|\phi_{+}\rangle-|000\rangle|\psi_{+}\rangle)$\\
$\i\sigma_{2}\otimes \sigma_{1}\otimes \sigma_{1}$ & $\frac{1}{2}(|111\rangle|\phi_{-}\rangle+|101\rangle|\psi_{-}\rangle-|011\rangle
|\phi_{+}\rangle-|000\rangle|\psi_{+}\rangle)$\\
$i\sigma_{2}\otimes i\sigma_{2}\otimes \sigma_{1}$ & $\frac{1}{2}(|111\rangle|\phi_{-}\rangle-|101\rangle|\psi_{-}\rangle-|011\rangle
|\phi_{+}\rangle+|000\rangle|\psi_{+}\rangle)$\\
$I\otimes\sigma_{1}\otimes \sigma_{1}$ & $\frac{1}{2}(|011\rangle|\phi_{-}\rangle+|001\rangle|\psi_{-}\rangle+|111\rangle|\phi_{+}\rangle
+|100\rangle|\psi_{+}\rangle)$\\
$I\otimes i \sigma_{2}\otimes \sigma_{1}$ & $ \frac{1}{2}(|011\rangle|\phi_{-}\rangle-|001\rangle|\psi_{-}\rangle+|111\rangle
|\phi_{+}\rangle-|100\rangle|\psi_{+}\rangle)$\\
$\sigma_{3}\otimes\sigma_{1}\otimes \sigma_{1}$ & $\frac{1}{2}(|000\rangle|\psi_{-}\rangle-|010\rangle|\phi_{-}\rangle+|111\rangle
|\phi_{+}\rangle-|100\rangle|\psi_{+}\rangle)$ \\
$\sigma_{3}\otimes i\sigma_{2}\otimes \sigma_{1}$ & $\frac{1}{2}(|000\rangle|\psi_{-}\rangle-|010\rangle|\phi_{-}\rangle
-|111\rangle|\phi_{+}\rangle+|100\rangle|\psi_{+}\rangle)$\\
$\sigma_{1}\otimes I \otimes \sigma_{1}$ & $\frac{1}{2}(|101\rangle|\phi_{-}\rangle+|111\rangle|\psi_{-}\rangle
+|001\rangle|\phi_{+}\rangle+|010\rangle|\psi_{+}\rangle)$\\
$\sigma_{1} \otimes \sigma_{3}\otimes \sigma_{1}$ & $\frac{1}{2}(|101\rangle|\phi_{-}\rangle-|111\rangle|\psi_{-}\rangle
+|001\rangle|\phi_{+}\rangle-|010\rangle|\psi_{+}\rangle)$\\
$i\sigma_{2}\otimes I\otimes \sigma_{1}$ & $\frac{1}{2}(|000\rangle|\phi_{+}\rangle-|100\rangle|\phi_{-}\rangle-|111\rangle|\psi_{-}\rangle
+|010\rangle|\psi_{+}\rangle)$\\
$i\sigma_{2}\otimes\sigma_{3}\otimes \sigma_{1}$ & $\frac{1}{2}(|101\rangle|\phi_{-}\rangle+|111\rangle|\psi_{-}\rangle-
|001\rangle|\phi_{+}\rangle-|010\rangle|\psi_{+}\rangle)$\\
\hline
\hline
\end{tabular}
\end{table}

The capacity of superdense coding is defined as,

\begin{equation}
X(\rho^{AB})=log_{2}d_{A}+S(\rho^{B})-S(\rho^{AB}),
\end{equation}

where $d_{A}$ is the dimension of Alice's system, $S(\rho)$ is
von-Neumann entropy. For the state $|\psi_{5}\rangle$,
$X(\rho^{AB})=3+2-0=5$. The Holevo bound of a multipartite quantum
state gives the maximum amount of classical information that can
be encoded \cite{Bruss}. It is equal to five, for the five-qubit
state ($log_{2}N$). Thus, the super dense coding reaches the
"Holevo bound" allowing five classical bits to be transmitted
through three quantum bits.

\section{Conclusion}

We have illustrated an example of a TMES for teleportation
of an unknown single and two qubit states. This state is also a
very useful resource for superdense coding. The superdense coding
capacity for the state reaches Holevo bound of five classical
bits. This also gives a picture, about what kind of lesser
entangled states could be useful for teleportation and superdense
coding. The study of the decoherence properties of this state and the 
Brown state also needs careful investigation in case of any practical application.
The usefulness of $|\psi_{5}\rangle$ , and the Brown state for many other
applications like the Quantum error correction and one way quantum computing 
needs extensive investigation. The physical realization of these states
in laboratory conditions is yet another challenge. 

\end{document}